\documentclass[aps,twocolumn,nofootinbib]{revtex4-1} 
\usepackage{pifont}
\usepackage{graphicx}
\usepackage{amssymb}
\usepackage{dcolumn}
\usepackage{amsmath}
\usepackage{bm}
\usepackage{textcomp}
\usepackage{epstopdf}
\usepackage{color}
\usepackage{ulem}
\usepackage[toc,page,title,titletoc,header]{appendix}
\usepackage[colorlinks,
linkcolor=blue,
anchorcolor=blue,
citecolor=blue
]{hyperref}
\usepackage{txfonts}
\usepackage{float}
\usepackage{multirow}
\begin{document}

\title{Controlling the RKKY interaction and heat transport in \\ a Kitaev spin liquid via $Z_2$ flux walls}
\author{Jun-Hui Zheng}
\author{Arne Brataas}
\affiliation{Center for Quantum Spintronics, Department of Physics, Norwegian University of Science and Technology, NO-7491 Trondheim, Norway}

\begin{abstract}
Kitaev spin liquids (KSLs) contain both itinerant Majorana fermions and localized $Z_2$ flux excitations. The mobile fermions transport energy, while the fluxes hinder heat flow. By controlling the flux configuration, one can separate a KSL into different domains via $Z_2$ flux walls. We show that at low temperatures, the wall significantly suppresses the Majorana-mediated Ruderman-Kittel-Kasuya-Yosida (RKKY) interaction and blocks the heat transfer between domains.
\end{abstract}

\maketitle

\section{Introduction}

Quantum spin liquids (QSLs) have attracted extensive attention because they lack long-range magnetic order while exhibiting long-range quantum entanglement \cite{Savary_2016}.
The concept of a `QSL' was developed for describing frustrated magnetism \cite{ANDERSON1973153} and understanding high-temperature superconductivity \cite{1987Anderson}. Moreover, QSLs are promising candidates for topological quantum computation governed by anyonic excitations \cite{Balents,Han2012,Ville2017}. Among the candidate models of QSLs, Kitaev’s honeycomb lattice model with anisotropic Ising-like exchange interactions is exactly solvable with rich anyon statistics \cite{kitaev2006}. Candidate materials of the Kitaev spin liquid (KSL) \cite{Jackeli2009,Singh2012,Takagi2019,Banerjee2016} have been observed in experiments through inelastic neutron scattering \cite{Knolle2014}, Raman scattering \cite{Knolle113,Nasu2016}, and thermal (Hall) transport measurements \cite{Nasu2017,Kasahara2018,Hentrich2019,Pidatella2019}.

In KSLs, the electron spin fractionalizes into two types of anyons that correspond to itinerant Majorana fermions and localized $Z_2$ fluxes \cite{kitaev2006,Baskaran2007,Knolle2014,Jansa2018}. Recent works have studied how the related fractionalization affects thermal transport and heat capacity in the equilibrium limit \cite{Nasu2015,Nasu2017,Pidatella2019}. Another recent numerical study demonstrates the possibility of nonlocal dynamic spin transport via Majorana fermions \cite{Minakawa2020}. In contrast to itinerant excitations, the $Z_2$ fluxes are localized and have received less attention since the flux configuration is conserved in the KSL. However, the same feature creates an opportunity for recording data in flux configurations due to their robustness. Theoretically, each unit lattice cell can store one-bit information by imprinting or removing a flux, representing an atomic-scale data entity and possibly enabling high-density storage. On the other hand, the flux configuration influences the transport properties of itinerant Majorana fermions, enabling detection of the configuration via transport measurements or nonlocal measurements.

\begin{figure}[t]
\centering
\includegraphics[width=0.8\columnwidth]{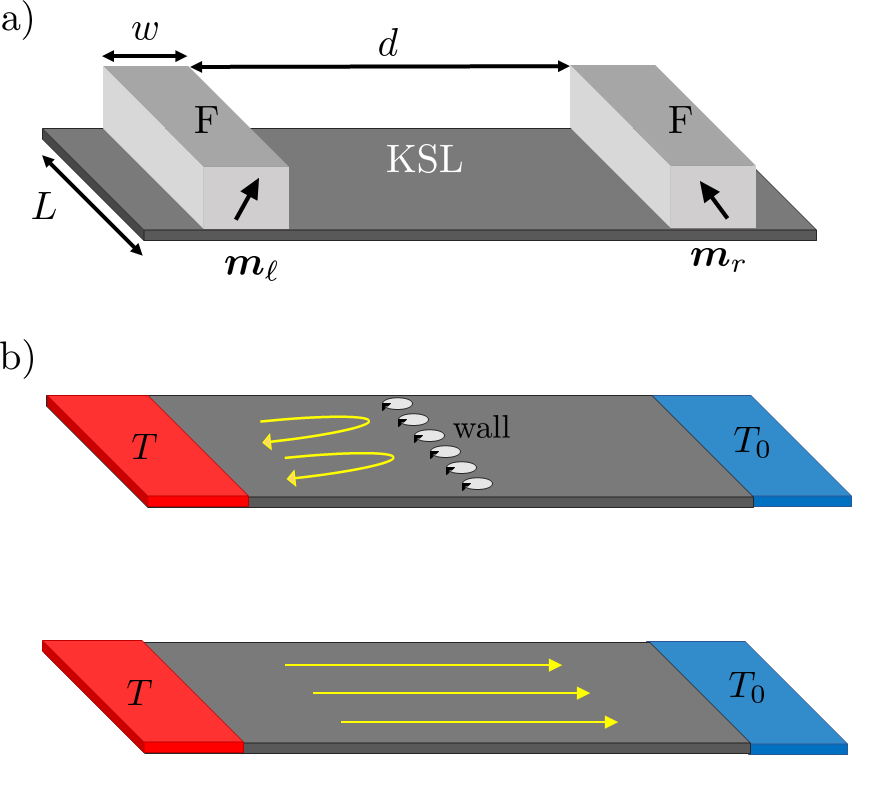}
\caption{a) A KSL is in contact with two ferromagnets (F). b) Heat transport in the presence and in the absence of a flux wall. The red part has a higher temperature than the blue part; $T>T_0$.}
\label{fig:nonlocal}
\end{figure}

\begin{figure}[t]
\centering
\includegraphics[width=0.8\columnwidth]{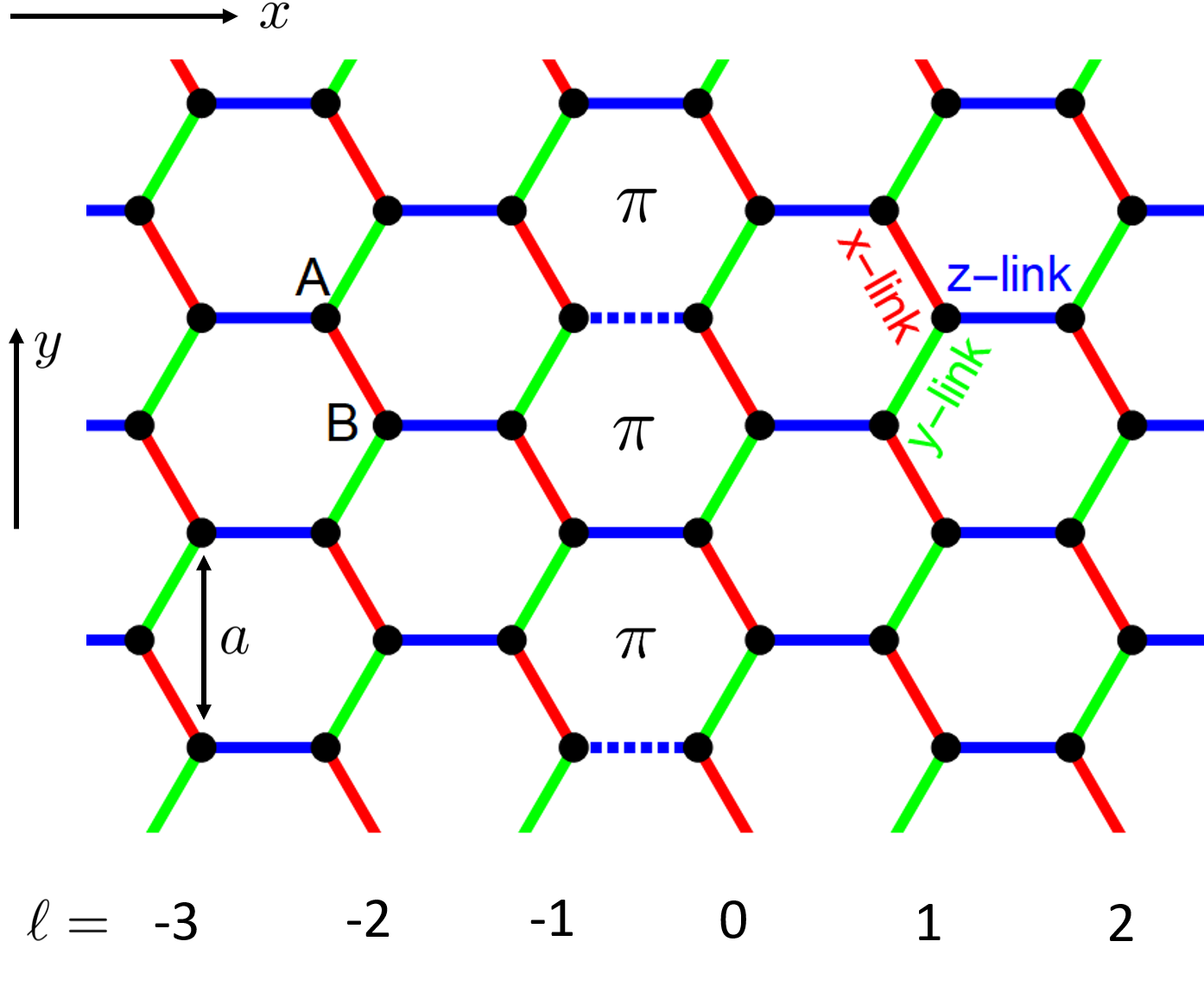}
\caption{A configuration of $u_j$ for a KSL with a column of fluxes. For the blue solid lines, $u_j=1$, and for the blue dashed lines, $u_j=-1$.}
\label{fig:config}
\end{figure}

This article considers two specific flux configurations with and without a $Z_2$ flux wall in the middle. The flux wall is composed of a column of $Z_2$ fluxes, as shown in Fig.\,\ref{fig:nonlocal}b and Fig.\,\ref{fig:config}; this is a metastable excited state and represents a bipartite separation of the system, similar to that by domain walls in magnetic materials. These excitations have a long lifetime since the flux configuration is conserved in the KSLs. In this article, we investigate how the wall affects the equilibrium exchange interaction mediated by KSLs and the dynamic heat transport in KSLs. The rest of article is organized as follows.

In Sect.\,\ref{hams}, we introduce the Majorana representation of the Kitaev model developed in Ref.\,\cite{Feng2007} and our conventions. In Sect.\,\ref{rkky}, as shown in Fig.\,\ref{fig:nonlocal}a, we develop the Majorana fermion-mediated Ruderman-Kittel-Kasuya-Yosida (RKKY) interaction \cite{Ruderman1954,Kasuya1956,Yosida1957,Legg2019} between ferromagnets. We demonstrate that the exchange interaction between two ferromagnets with magnetic moment densities $\bm m_\ell$ and $\bm m_r$ is a linear combination of $(m_\ell^\alpha)^2 (m_r^\beta)^2$, where $\alpha, \beta = x,y,z$ instead of the $\bm m_\ell \cdot \bm m_r$ in metal-mediated interactions \cite{Brataas2005}. The corresponding nonlocal torque depends on both magnetic moment densities. We also show that a flux wall strongly suppresses the exchange interaction. In Sect.\,\ref{heatt}, we consider the influence of flux walls on heat transport (Fig.\,\ref{fig:nonlocal}b). We reveal that the transport properties strongly differ for a KSL in the two flux configurations. When the wall is present, the system separates into two domains, and the wall blocks the heat exchange between domains. Specifically, the low-energy itinerant Majorana particles cannot pass through the wall and are completely backscattered. This feature allows us to differentiate the two configurations in thermal transport measurements. In Sect.\,\ref{creat}, we discuss how to experimentally create fluxes and flux walls in the KSL. In Sect.\,\ref{concl}, we give a short summary with an outlook.

\section{Hamiltonian}\label{hams}

As shown in Ref.\,\cite{Feng2007}, representing the spin operators in terms of Majorana fermions enables the analytical solving of the Kitaev model. In this representation,
the Hamiltonian reads
\begin{equation}\label{hamw}
\hat{H}_{0} =  i \sum_{j\in A}( J_x \hat c_j \hat c_{j_x} + J_y \hat c_j \hat c_{j_y} +  J_z \hat{u}_j \hat c_j \hat c_{j_z}),
\end{equation}
where, as shown in Fig.\,\ref{fig:config}, the index $j \in A$ denotes a site of the A-sublattice on a honeycomb lattice, $j_\alpha$, where $\alpha = x,y,z$, refers to the adjacent B-site via an $\alpha$-link, $\hat{u}_{j} = - i \hat b_j \hat b_{j_z}$ is a link variable on a $z$-link, and $\hat c$ and $\hat b$ are Majorana operators.

The operator $\hat c$ represents itinerant Majorana particles, while $\hat b$ is related to localized $Z_2$ fluxes. The eigenvalues associated with the link variables are ${u}_{j} = \pm 1$, and they are good quantum numbers \cite{Feng2007}. When a hexagonal cell has two opposite quantum numbers ${u}_{j}$ on its edges, a $\pi$ flux exists in the cell. Throughout the article, we focus on the case $J_\alpha =J$ for $\alpha=x,y,z$. In this case, the ground state has no fluxes and has a large degeneracy \cite{Feng2007}. In addition, the itinerant Majorana excitations are gapless \cite{kitaev2006,Pedrocchi,Baskaran2007}. On the other hand, a single flux excitation has a finite energy $\Delta \simeq 0.1536 J$ \cite{kitaev2006}. In the following, we focus on the low-temperature case, $T \ll \Delta$, where the flux excitations are strongly suppressed in thermal equilibrium.

\section{ Nonlocal RKKY exchange interaction} \label{rkky}
In this section, we study the KSL-mediated exchange interaction between magnets. As shown in Fig.\,\ref{fig:nonlocal}a, the KSL contacts two ferromagnets with magnetic moment densities $\bm m_\ell$ and $\bm m _r$. The exchange interaction between the KSL and the ferromagnets is assumed to be $\hat V_{\kappa} = -\lambda \sum_{q}  { \bm m}_\kappa \cdot \bm  {\hat \sigma}_q$, where $\kappa = \ell, r$ and $q$ runs over sites on each interface.

We assume that the magnetization $\bm m_\kappa$ is static and classically treat $\bm m_\kappa$. Moreover, the exchange interaction $\hat V_{\kappa}$ is assumed to be weak so that the KSL is perturbed, i.e., $\lambda |\bm m_\kappa| \ll J_\alpha$. As a result, we find that when the two magnets have a finite separation $d$, the nonlocal contribution to the free energy of the full system is
\begin{equation}\label{free_nonloc}
F_{\text{N}}(\bm m_\ell, \bm m_r, d) = \eta(d) L w^2 \sum_{\alpha\beta} (m_\ell^\alpha)^2 \mathcal{N}_{\alpha\beta} (m_r^\beta)^2,
\end{equation}
where the sum $\alpha,\beta$ includes $x,y$ and $z$, $\mathcal{N}$ is a $3\times 3$ matrix independent of the separation $d$, and $\eta(d)$ depends on the distance $d$ and vanishes when $d\rightarrow \infty$. $L$ and $w$ are geometric parameters of the magnets. From this free energy, we find that the nonlocal toque on the $\kappa$ side, $\bm \tau_\kappa = -\nabla_{\bm m_\kappa} F_{\text{N}}$, depends on both magnetic moment densities.

Below, we derive Eq.\ \eqref{free_nonloc} in two steps. Note that the spin operator $\hat \sigma_j^\alpha$ creates excitations with a pair of neighboring fluxes. The lowest energy in these excitations is approximately $0.267 J$ \cite{kitaev2006}, which is much larger than the temperature we consider. Thus, the interaction $\hat{V}_\kappa$ couples high energy excitations. In the first step, we develop a low-energy effective interaction $\hat{V}^\text{eff}_\kappa$ between the ferromagnets and the KSL in the flux-free sector.

Following Kitaev's discussion, the low-energy effective interaction, up to the second order in $\hat V_\kappa$, can be written as $\hat{V}_\kappa^\text{eff}= \Pi_0 \hat V_\kappa G'(E_0) \hat V_\kappa \Pi_0$, where $\Pi_0$ is the projector onto the flux-free sector and $G'$ is the Green's function with the flux-free sector excluded \cite{kitaev2006}. Formally, the low-energy effective interaction in the flux-free sector can be written as
\begin{equation}\label{veff}
\hat{V}_\kappa^\text{eff}= \mathcal{E}(\bm m_\kappa) + \sum_{j,\alpha}  {\lambda}^\text{eff}_{\alpha} (\bm m_\kappa) \hat \sigma_j^\alpha  \hat\sigma_{j_\alpha}^\alpha,  %
\end{equation}
where the first term contributes an energy shift and the second term corrects exchange interactions between neighbors.

We determine the parameters in Eq.~\eqref{veff} by requiring the free energy to satisfy
\begin{equation}
\mathcal{F}(\hat{H}_0 +  \hat V_\kappa^\text{eff}) = \mathcal{F}(\hat{H}_0 + \hat V_\kappa)
\end{equation}
where $\mathcal{F}(\hat{h}) \equiv - k_B T \ln \text{Tr}[\exp{(-\beta \hat{h})} ]$ with $\beta =1/k_B T$. Order by order expanding the two functions $\mathcal{F}$s and comparing them term by term, we have
\begin{eqnarray}
 && \mathcal{E}(m_\kappa^\alpha) = -\frac{\lambda^2}{2}\sum_{q,\alpha} (m_\kappa^\alpha)^2 \mathcal{U}_{qq}^\alpha,\\
 &&   {\lambda}^\text{eff}_{\alpha}(m_\kappa^\alpha) = -\frac{\lambda^2 (m_\kappa^\alpha)^2 \mathcal{U}_{j j_\alpha}^\alpha }{\langle \hat \sigma_j^\alpha  \hat\sigma_{j_\alpha}^\alpha\rangle},
\end{eqnarray}
where $\mathcal{U}_{qq'}^\alpha =  \int_0^\beta d \tau  \langle T_\tau \sigma_{q}^\alpha(\tau) \sigma_{q'}^\alpha (0)  \rangle_c$. Throughout, the expectation value $\langle  \cdot \rangle_c $ means a connected Feynman diagram, which is evaluated in the absence of ferromagnets. The details can be found in Appendix A.

In the second step, we develop the RKKY interaction between the two ferromagnets. Note that the total low-energy effective Hamiltonian becomes $\hat{H} =\hat{H}_0 +\hat{V}_\ell^\text{eff}+ \hat{V}_r^\text{eff}$. To find the effective potential describing the coupling between the two ferromagnets, we calculate the full free energy $\mathcal{F}(\hat{H})$.
The leading order of the nonlocal part of the free energy is then proportional to $\lambda_{\alpha}^\text{eff}(\bm m_\ell)\lambda_{\beta}^\text{eff}(\bm m_r)$. Expressing this term in the form of Eq.~\eqref{free_nonloc}, we finally obtain
\begin{equation}\label{free3}
\eta(d) \mathcal{N}_{\alpha\beta} =  -\lambda^4 \mathcal{Z}_\alpha \mathcal{Z}_\beta  \sum_{j,j'} {\chi}_{jj'}^{\alpha\beta},
\end{equation}
where $ \mathcal{Z}_\alpha = \mathcal{U}_{j j_\alpha}^\alpha /  \langle \hat \sigma_j^\alpha  \hat\sigma_{j_\alpha}^\alpha\rangle_c  $ is spatially independent, the indices $j$ and $j'$ run over A sites on the $\ell$-side and $r$-side interfaces, respectively, and ${\chi}_{jj'}^{\alpha\beta} $ is a nonlocal correlation, where
\begin{equation}\label{chi}
 {\chi}_{jj'}^{\alpha\beta} = \int_0^\beta d\tau \langle \hat T_\tau   \hat \sigma_j^\alpha(\tau)  \hat\sigma_{j_\alpha}^\alpha(\tau)  \hat \sigma_{j'}^\beta (0) \hat\sigma_{j'_\beta}^\beta (0)   \rangle_c.
\end{equation}

\begin{figure}[t]
\centering
\includegraphics[width=0.8\columnwidth]{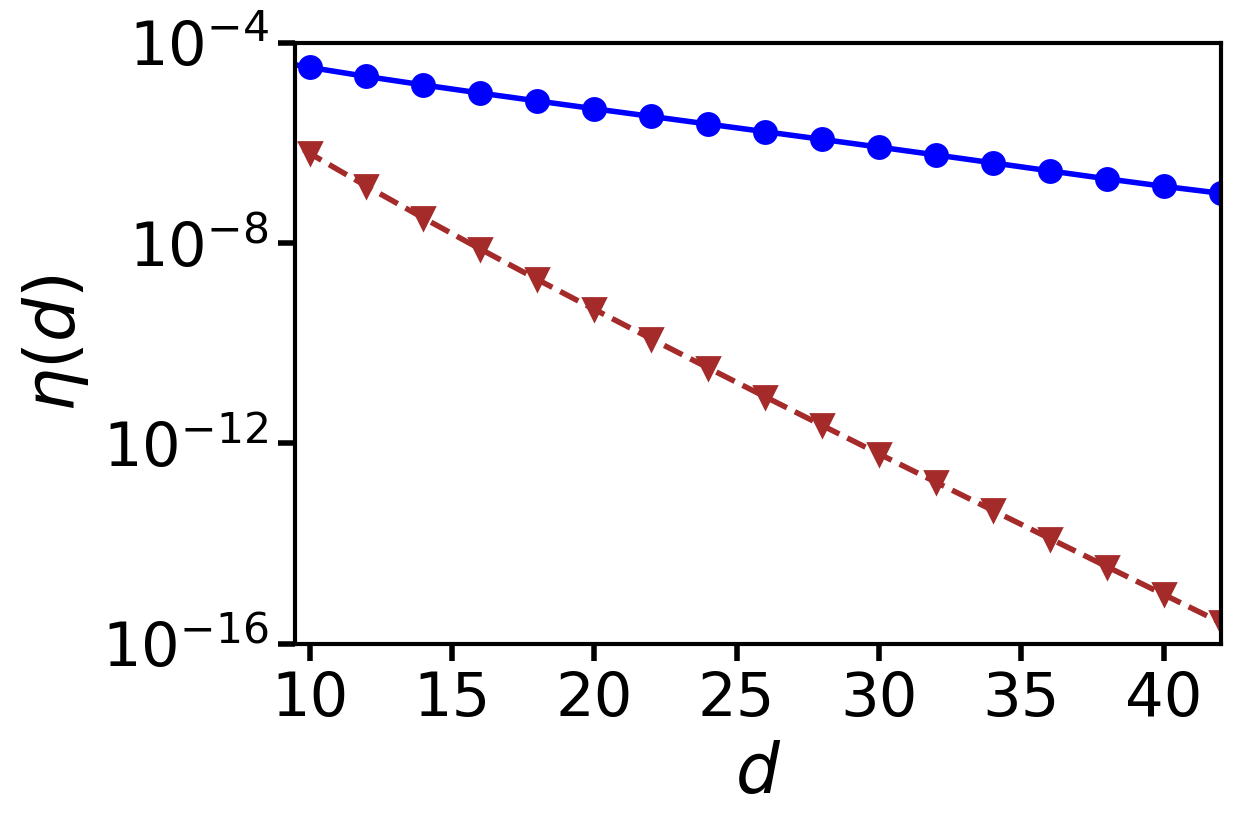}
\caption{The function $\eta$ as a function of the separation $d$ in the absence (blue curve) and presence (red curve) of a flux wall. We assume that the width of the ferromagnets $w$ is much smaller than the separation between the ferromagnets $d$. $d$ is in units of $\sqrt{3}a/2$, where $a$ is the lattice constant, as shown in Fig.\,\ref{fig:config}.}
\label{fig:exchange}
\end{figure}

According to the parity symmetry of the system, the matrix $\mathcal{N}$ has the form $\mathcal{N}_{xx}=\mathcal{N}_{yy}= n_1$, $\mathcal{N}_{z z} =n_2$, $\mathcal{N}_{xy}= \mathcal{N}_{yx}= n_3$, $\mathcal{N}_{\alpha z} = \mathcal{N}_{z \alpha} =n_4$, where $\alpha = x,y$. When the two magnets are well separated ($d>10\sqrt{3}a/2$, where $a$ is the lattice constant), $\bm{n} =(n_1,n_2,n_3,n_4)$ becomes almost spatially independent. Numerically, we find $ \bm n= (-1, -3.98,-1,2)$ in the absence of a flux wall and $ \bm n= (-1, -2.08,-1,1.56)$ with a flux wall in the middle, as shown in Fig.\,\ref{fig:nonlocal}b. The coefficient $\eta$ exponentially decays with respect to the separation $d$, as shown in Fig.\,\ref{fig:exchange}. In the plot, we set $\lambda^2 \mathcal{Z}_\alpha =1$. We find that a flux wall strongly suppresses the magnitude of $\eta$, showing that flux walls effectively decouple the ferromagnets. More details on the nonlocal correlation can be found in Appendix B.

\section{Influence of a flux wall on the heat transport} \label{heatt}

References \cite{Nasu2015,Nasu2017,Pidatella2019} explored the heat capacity $C_v$ and heat conductivity $\kappa$ without flux walls. In the low-temperature limit, the conductivity is proportional to $T$, similar to the conductivity in disordered graphene. In this section, we consider how flux walls change heat transport.

Since different $u$-configurations (i.e., different gauge representations) for the same spatial flux distribution are equivalent under a unitary transformation \cite{Feng2007}, we focus on one $u$-configuration. The simplest $u$-configuration with a wall is shown in Fig.\,\ref{fig:config}. The link variables $u_j$ in the middle column periodically change from $-1$ to $1$. Using the Fourier transformation along the $y$-direction, we find that the Hamiltonian \eqref{hamw} with a wall separates into contributions from the left domain, right domain, and wall
\begin{equation}\label{hw}
    \hat{H}_{w} = \sum_{k_y \in [ -\pi,0]} [\hat{h}^L_{k_y} + \hat{h}^R_{k_y} + \hat{V}_{k_y}],
\end{equation}
where
\begin{eqnarray}\label{hlrv}
 \hat{h}^L_{k_y} &=&  \sum_{l}  \Big[-4J\cos \frac{k_y}{2} \hat c^\dagger_{l,k_y} \sigma_y  \hat c_{l,k_y} - 2J ( i\hat c^\dagger_{l-1,k_y,B}  \hat c_{l,k_y,A} +h.c.) 
 \Big]\notag\\
\hat{h}^R_{k_y} &=& \sum_{l\geq 0} \Big[ -4J \cos\frac{k_y}{2} \hat c^\dagger_{l,k_y} \sigma_y  \hat c_{l,k_y}- 2J ( i\hat c^\dagger_{l,k_y,B}  \hat c_{l+1,k_y,A} +h.c.) 
 \Big]\notag\\
\hat{V}_{k_y}  &=& - 2J  (i\hat c^\dagger_{-1,k_y,B}  \hat c_{0,k_y+\pi/a,A} +h.c.).
\end{eqnarray}
Here, $\hat c^\dagger_{l,k_y} = \big(\hat c^\dagger_{l,k_y,A},\hat c^\dagger_{l,k_y,B} \big)$ is a two component operator in the space of the sublattices $A$ and $B$ and $l$ refers to the column of zigzag lines, as shown in Fig.\,\ref{fig:config}. In Eq. \eqref{hw}, we use the fact that for a Majorana operator $\hat{c}_j$, its Fourier transform satisfies $\hat c^\dagger_{l,k_y, s} = \hat c_{l,-k_y,s}$
for $s =A,B$.

Note that the term $\hat{V}_{k_y}$ couples the left domain and the right domain by a momentum shift $\Delta k_y = \pi/a$. This allows us to resort the terms in Eq.~\eqref{hw}. Consequently, the Hamiltonian in $(x, k_y)$ space becomes a combination of one-dimensional models with different $k_y$,
\begin{equation}
    \hat{H}_{w} = \sum_{k_y \in [-\pi, 0]} \hat{h}_{k_y},
\end{equation}
where
\begin{equation}
 \hat{h}_{k_y} = \hat{h}^L_{k_y} + \hat{h}^R_{k_y+\pi/a} + \hat{V}_{k_y}.
\end{equation}
In the following, we illustrate that the tunneling of the Majorana particle through the flux wall is forbidden at low energy. Consequently, the low-energy itinerant Majorana particles are trapped in one domain, and the heat exchange between the domains is blocked.

For different momenta $k_y$, the one-dimensional Hamiltonian $\hat{h}_{k_y}$ decouple from each other; therefore, the two-dimensional tunneling problem of $\hat{H}_{w} $ reduces to the one-dimensional scattering problem of $\hat{h}_{k_y}$. For the one-dimensional model $\hat h_{k_y}$, we consider an incident plane wave from the left with momentum $k_x$, in which case $\hat c_{l,k_y,A} \sim \hat c_{k_x,k_y,A} \exp{[i k_x l \sqrt{3} a/2]}$. From the formula $\hat{h}^L_{k_y}$ shown in Eq. \eqref{hlrv}, we obtain the energy of the incident wave,
\begin{equation}\label{spec}
E(\bm k) = \pm 2J \sqrt{4\cos^2\frac{k_y a}{2}+4\cos\frac{\sqrt{3}k_x a}{2}\cos\frac{k_y a}{2}+1}.
\end{equation}
Below, we focus on the branch with positive energy. The negative branch has a similar result.
We are interested in the behavior of low-energy particles near the Dirac point. For those states near the Dirac cone $\bm k_0 =(2\sqrt{3}\pi/3a,-2\pi/3a)$, the spectrum becomes $E(\bm k)= \sqrt{3}J a | \bm k -\bm k_0|$. The right moving particle is scattered when it meets the flux wall. If the particle passes through the wall with a finite possibility, then in the right domain away from the wall, there should exist an asymptotic state with energy $ E(\bm k)$ for some momentum $(k'_x, k_y+\pi/a)$. However, from Eq.\,\eqref{spec}, we find that for any given momentum $k'_x$, we have
\begin{equation}
 E(k'_x, k_y+\pi/a) \geq 2J[-2\sin(k_y a/2)-1] > \Delta,
\end{equation}
for $k_y$ near to $k_0^y$.
Consequently, a low-energy incident state with transverse momentum $k_y$ arising from the left domain has no corresponding low-energy states with transverse momenta $k_y+\pi/a$ in the right domain. As a result, all low-energy incident states from the left decay in the right domain and are completely backscattered.

We can estimate the penetration depth of the incident wavefunctions in the right domain. Since the wavefunction decays on the right, the wavenumber $k'_x$ becomes a complex number with Im$[k'_x]>0$. For instance, for an incident wave with $\bm k=\bm k_0$, the condition $E(k'_x, k_y+\pi/a)= E(\bm k_0) =0$ gives Im$[k'_x]\simeq 0.63/a$. As a result, the penetration depth of the incident wave is approximately $1/\text{Im}[k'_x] \simeq 1.6 a$. The wavefunction decays quickly in the right domain.

We have discussed a wall constructed by a single column of flux. Interestingly, not all wall configurations efficiently forbid tunneling. Since a column of fluxes changes the momentum $k_y$ by $\pi/a$, when the width becomes twice as large, the momentum $k_y$ changes twice and returns to its initial value. In this case, the particle can partially pass through the wall again. Consequently, only walls with an odd number of columns of fluxes completely block heat transfer between the two domains. These properties of the flux wall can be measured in heat transport experiments, as shown in Fig.\,\ref{fig:nonlocal} b). The two configurations with and without a flux wall can be distinguished experimentally.

\section{Creating fluxes} \label{creat}

\begin{figure}[t]
\centering
\includegraphics[width=0.8\columnwidth]{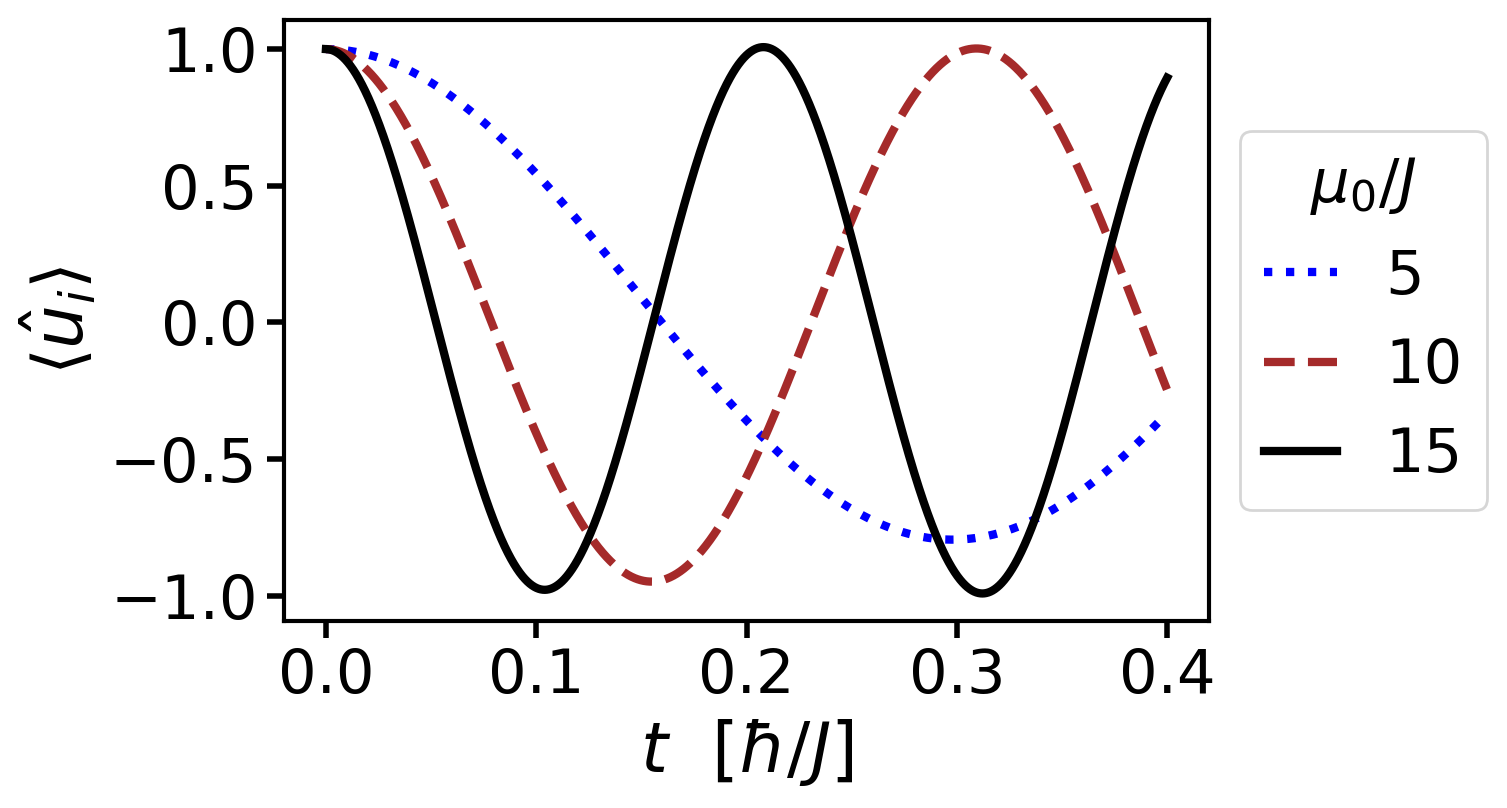}
\caption{The temporal evolution of the expectation value of the link variable. A pair of vortices are created when $\langle \hat u_i \rangle$ changes from $1$ to $-1$ and removed when it recovers to $1$.}
\label{fig:result1}
\end{figure}

In this section, we discuss how to create fluxes in the KSL. Conceptually, fluxes can be excited by a non-Kitaev type of exchange interaction, for instance, the Heisenberg type interaction \cite{Knolle113,Perreault2016}. Here, we propose the use of a ferromagnetic tip to magnetize the local spin $\bm \sigma_i$ at site $i \in A$. The interaction between the tip and the local spin is supposed to be ferromagnetic, $ \hat{V} = - \lambda' \bm \hat{\sigma}_i \cdot \bm M$. We assume that the magnetic moment on the tip is giant and treat it classically. We
also assume that the magnetic moment runs along the $z$-direction. Consequently, we obtain $
\hat{V} = - \mu_0 \hat \sigma_i^z = i \mu_0 \hat{c}_i \hat{b}_i$,
where $\mu_0 = \lambda'M_z > 0$. This interaction changes both the number of
itinerant Majorana particles and that of fluxes.

We calculate the expectation value of $\hat u_i$ during the dynamic evolution after coupling with the tip. We start from the initial ground state $\hat{u}_q =1$. When $\langle \hat u_i \rangle$ changes to $-1$, a pair of fluxes are created in the two adjacent honeycomb cells of $\hat u_i$. In the numerical calculation, we apply the Hartree-Fock mean-field approximation to decouple the term $i J_z \hat{u}_i \hat c_i \hat c_{i_z}$ \cite{Minakawa2020}.
In Fig.\,\ref{fig:result1}, we show the numerical result of the temporal evolution of the link variable $\langle\hat u_i \rangle$.
The result shows that it is possible to create fluxes when $\mu_0$ is sufficiently large ($\mu_0 \sim 10 J$). In creating a pattern of fluxes or a flux wall, a series of operations of the local magnetization should be implemented. The time for creating a new flux pair depends on the existing flux distribution because of the flux-flux interaction. This increases the difficulty of creating a general pattern of fluxes.

\section{Conclusion} \label{concl}

In conclusion, we developed a low-energy effective theory to describe the equilibrium nonlocal exchange interactions mediated by KSLs in the presence and absence of a flux wall and to calculate the transport of Majorana particles though the flux wall. We found that a flux wall significantly suppresses the equilibrium RKKY interaction between magnetic moments attached to the KSL. In addition, the flux wall significantly changes the heat transport. The flux wall completely blocks heat transfer in the low-temperature case. Heat transport measurements can detect the presence of the wall. We also discussed how to create fluxes in a KSL. Our study opens opportunities for storing and reading data in a KSL by using two simple flux configurations. It is interesting to see that a KSL can store data even though it lacks magnetic order.

This work was supported by the Research Council of Norway through its Centres of Excellence funding scheme (project no.\,262633, ``QuSpin'').

\onecolumngrid
\appendix
\section{The coefficients of the effective interaction}
In this section, we derive the coefficients of the effective interaction in Eq. (3). For this purpose, we assume that all coefficients in the interaction $\hat{V}_\kappa$ depend on the position, $\lambda {m}_\kappa^\alpha \rightarrow \lambda_{q}^\alpha$, so that $\hat V_{\kappa}  = - \sum_{q\alpha}  \lambda_{q}^\alpha   {\hat \sigma}_q^\alpha$.
From the equation $\hat{V}_\kappa^\text{eff}= \Pi_0 \hat V_\kappa G'(E_0) \hat V_\kappa \Pi_0$, we obtain
\begin{equation}\label{sm_veff}
\hat{V}_\kappa^\text{eff}= \sum_{q,\alpha} \epsilon_{q\alpha} + \sum_{j,\alpha}  {\lambda}^\text{eff}_{j\alpha} \hat \sigma_j^\alpha  \hat\sigma_{j_\alpha}^\alpha,  %
\end{equation}
where $\epsilon_{q\alpha} \propto {\lambda}_{q}^\alpha {\lambda}_{q}^\alpha$ and  ${\lambda}^\text{eff}_{j\alpha} \propto {\lambda}_{j}^\alpha {\lambda}_{j_\alpha}^\alpha$. Expanding the free energy
$\mathcal{F}(\hat{H}_0 +  \hat V_\kappa^\text{eff})$ with respect to $\epsilon_{q\alpha}$ and ${\lambda}^\text{eff}_{j\alpha}$, where $\mathcal{F}(\hat{H}) \equiv - k_B T \ln \text{Tr}[\exp{(-\beta \hat{H})} ]$, we obtain
\begin{equation}
    \mathcal{F}(\hat{H}_0 +  \hat V_\kappa^\text{eff})  =  \mathcal{F}(\hat{H}_0) + \sum_{q\alpha}\epsilon_{q\alpha} + \sum_{j\alpha} \langle \sigma_j^\alpha  \hat\sigma_{j_\alpha}^\alpha  \rangle {\lambda}^\text{eff}_{j\alpha}.
\end{equation}
Here, we use the fact that the free energy is the generating functional for connected Green's functions. On the other hand, expanding the free energy
$\mathcal{F}(\hat{H}_0 +  \hat V_\kappa)$ with respect to  ${\lambda}_q^{\alpha}$, we have
\begin{equation}
    \mathcal{F}(\hat{H}_0 +  \hat V_\kappa)  =  \mathcal{F}(\hat{H}_0) - 
 \sum_{q\alpha}  \lambda_{q}^\alpha \langle  {\hat \sigma}_q^\alpha\rangle   
    -  \frac{1}{2}\sum_{q\alpha,q'\alpha'} \int_0^\beta d\tau \langle T_\tau\sigma_q^\alpha(\tau) \hat\sigma_{q'}^{\alpha'}(0) \rangle_c {\lambda}_{q}^\alpha{\lambda}_{q'}^{\alpha'}.
\end{equation}
Note that $\langle{\hat \sigma}_q^\alpha\rangle=0$ and $\langle T_\tau\sigma_q^\alpha(\tau) \hat\sigma_{q'}^{\alpha'}(0) \rangle$ are finite only when $\alpha=\alpha'$ and $q =q'$ or when $\alpha=\alpha'$ and $q$ and $q'$ are nearest neighbors. Therefore, we obtain
\begin{equation}
    \epsilon_{q\alpha} = - \frac{1}{2} {\lambda}_{q}^\alpha {\lambda}_{q}^\alpha  \int_0^\beta d\tau \langle T_\tau\sigma_q^\alpha(\tau) \hat\sigma_{q}^{\alpha}(0) \rangle_c,~~~~~   {\lambda}^\text{eff}_{j\alpha} = - {\lambda}_{j}^\alpha {\lambda}_{j_\alpha}^\alpha 
    \int_0^\beta d\tau \langle T_\tau\sigma_j^\alpha(\tau) \hat\sigma_{j_\alpha}^{\alpha}(0)\rangle_c/\langle \sigma_j^\alpha  \hat\sigma_{j_\alpha}^\alpha  \rangle.
\end{equation}

\subsection{Details for the expansion of the free energy}
Here is an illustration of how the free energy is expanded. Let us consider the free energy $\mathcal{F}(\hat{H}_0 +  \hat V_\kappa)$. In the language of Feyman's path integral,
\begin{equation}
    \mathcal{F}(\hat{H}_0 +  \hat V_\kappa) = -k_B T \ln \int  \prod_i d b_i d c_i \exp\left\{{-S_0(b, c)-\int d\tau  \sum_{q\alpha}  \lambda_{q}^\alpha   {\sigma}_q^\alpha(\tau) }\right\}
\end{equation}
where $b$ and $c$ are Grassmann numbers corresponding to $\hat{b}$ and $\hat{c}$ and the action is $S_0 = -\sum_q \frac{1}{2}(b_q \partial_\tau b_q + c_q \partial_\tau c_q) + H_0(b,c)$. The spin $\sigma$ is a function of $b$ and $c$. Using Taylor expansion,
we directly obtain
\begin{equation}
    \mathcal{F}(\hat{H}_0 +  \hat V_\kappa)  = 
  \sum_{n}\frac{1}{n!} \sum_{\{q,\alpha\}_n}  \frac{\delta^n \mathcal{F}}{\delta  \lambda_{q_1}^{\alpha_1} \delta \lambda_{q_2}^{\alpha_2}
  \cdots \delta \lambda_{q_n}^{\alpha_n}}\lambda_{q_1}^{\alpha_1}  \lambda_{q_2}^{\alpha_2}
  \cdots  \lambda_{q_n}^{\alpha_n}.
\end{equation}
Note that ${\delta^n \mathcal{F}}/{\delta  \lambda_{q_1}^{\alpha_1} \delta \lambda_{q_2}^{\alpha_2}
\cdots \delta \lambda_{q_2}^{\alpha_n}}$ is the connected Green's function. For example,
\begin{eqnarray}
    \frac{\delta^n \mathcal{F}}{\delta  \lambda_{q_1}^{\alpha_1} \delta \lambda_{q_2}^{\alpha_2}} &=& (-k_B T) \frac{ \int  \prod_i d b_i d c_i [\int d\tau_1 d\tau_2{\hat \sigma}_{q_1}^{\alpha_1}(\tau_1) {\hat \sigma}_{q_2}^{\alpha_2}(\tau_2)] \exp\left\{-S_0(b, c)\right\}}{\int  \prod_i d b_i d c_i \exp\left\{{-S_0(b, c)}\right\}} \notag\\
    &&
    +k_B T \frac{ \int  \prod_i d b_i d c_i [\int d\tau_1 {\hat \sigma}_{q_1}^{\alpha_1}(\tau_1)] \exp\left\{{-S_0(b, c)}\right\}}{\int  \prod_i d b_i d c_i \exp\left\{{-S_0(b, c)}\right\}}\frac{ \int  \prod_i d b_i d c_i [\int d\tau_2{\hat \sigma}_{q_2}^{\alpha_2}(\tau_2)] \exp\left\{{-S_0(b, c) }\right\}}{\int  \prod_i d b_i d c_i \exp\left\{{-S_0(b, c)}\right\}} \notag\\
    &=& -\int d\tau [\langle T_\tau {\hat \sigma}_{q_1}^{\alpha_1}(\tau) {\hat \sigma}_{q_2}^{\alpha_2}(0) \rangle - \langle  {\hat \sigma}_{q_1}^{\alpha_1}(\tau)\rangle \langle{\hat \sigma}_{q_2}^{\alpha_2}(0) \rangle] \notag\\
    &=& -\int d\tau \langle T_\tau {\hat \sigma}_{q_1}^{\alpha_1}(\tau) {\hat \sigma}_{q_2}^{\alpha_2}(0) \rangle_c
\end{eqnarray}
is the connected Green's function.

\section{Nonlocal correlation ${\chi}_{jj'}^{\alpha\beta}$}
Next, we calculate the nonlocal correlation ${\chi}_{jj'}^{\alpha\beta}$.
The single-particle Green's function of the Majorana particle is $\mathcal{G}_{lm} = - \langle T_\tau \hat  c_l (\tau) \hat c_m(0)\rangle_c $. Below, we denote the Hamiltonian of the KSL as $\hat{H}_{0}= (1/4)\sum_{lm}  \hat{c}_l \mathcal{H}_{lm} \hat{c}_m $, where $\mathcal{H}$ is an antisymmetric matrix representation. Using the Heisenberg equation and the communication relation $\{\hat  c_l, \hat  c_m \} = 2 \delta_{lm}$, we obtain
\begin{equation}
  \mathcal{G}_{lm}(i\omega_n)  = \sum_{q} \frac{\mathcal{A}^q_{lm}}{i\omega_n - \varepsilon_q}
\end{equation}
where $\mathcal{A}^q_{lm} = 2 \phi_q(l) \phi_q^*(m)$, with $\phi_{q}$ being an eigenvector of $\mathcal{H}$ corresponding to the eigenvalues $\varepsilon_q$ and $\omega_n = (2n+1)\pi k_B T $.
By using the Majorana representation and the Wick theorem, we finally have
\begin{equation}
 {\chi}_{jj'}^{\alpha\beta}  = \sum_{q,q'} \frac{\big(\mathcal{A}^q_{jj'}\mathcal{A}^{q'}_{j'_\beta j_\alpha}  -\mathcal{A}^q_{jj'_\beta}\mathcal{A}^{q'}_{j' j_\alpha}  \big) \big[n_F(\varepsilon_q)-n_F(\varepsilon_{q'}) \big]}{  \varepsilon_q - \varepsilon_{q'}},
\end{equation}
where $n_F$ is the Fermi-Dirac distribution.

\twocolumngrid

\end{document}